\documentclass[12pt]{article}
\usepackage{amssymb}
\usepackage{amsmath}
\usepackage{graphicx}
\usepackage{indentfirst}
\usepackage{cite}

\begin{document}

\title{Free quarks and antiquarks versus hadronic matter}
\author{Xiao-Ming X$\rm{u}^1$ and Ru Pen$\rm{g}^2$}
\date{}
\maketitle \vspace{-1cm} 
\centerline{$^1$Department of Physics, Shanghai
University, Baoshan, Shanghai 200444, China}
\centerline{$^2$Department of Physics, Wuhan University of Science and 
Technology, Wuhan 430081, China} 
\begin{abstract}
Meson-meson reactions $A(q_1\bar {q}_1)+B(q_2\bar {q}_2) \to 
q_1+\bar {q}_1+q_2 +\bar {q}_2$ in high-temperature
hadronic matter are found to produce an
appreciable amount of quarks and antiquarks freely moving in hadronic matter
and to establish a new mechanism for deconfinement of quarks and antiquarks in
hadronic matter.
\end{abstract}



\newpage

The picture for hadronic matter is that this matter only consists of hadrons
and the evolution of this matter is determined by hadron scatterings and
hadron flows. Meson-meson elastic scatterings establish thermal states of
hadronic matter.
At the critical temperature of QCD phase transition quarks and antiquarks
automatically move from the small hadron volume toward the large volume of 
hadronic matter.
However, in the present work we show that this pure hadronic matter is mixed 
with free quarks and antiquarks. The reason is that meson-meson scatterings
into free quarks and free antiquarks at high temperature can 
cause the ratio of free-quark number density to number density of hadronic
matter to be
larger than 0.1 in a short time period of 0.5 fm/$c$. We also show that the 
reactions offer a new way for deconfinement near the critical temperature, and
name the deconfinement via the reactions collisional deconfinement. We 
address the occurrence of deconfinement from variation of number densities of
hadrons and free quarks and antiquarks in contrast to the variation of energy
density, chiral condensate and the screening mass of a heavy quark-antiquark
pair studied by lattice QCD \cite {Karsch}. 
We focus on the reactions of $\pi$, $\rho$,
$K$ and $K^\ast$ because the four hadron species are dominant meson species in
hadronic matter at RHIC \cite {adler,adams2,bearden}.

The cross section for the meson-meson reaction
$A(q_1\bar {q}_1)+B(q_2\bar {q}_2) \to 
q_1+\bar {q}_1+q_2 +\bar {q}_2$ is
\begin{eqnarray}
\sigma & = & \frac {(2\pi)^4}
                   {4\sqrt {(P_A \cdot P_B)^2 -m_A^2m_B^2}}    
    \int \frac {d^3p_{q_1^\prime}}{(2\pi)^32E_{q_1^\prime}}
         \frac {d^3p_{\bar {q}_1^\prime}}{(2\pi)^32E_{\bar {q}_1^\prime}}
         \frac {d^3p_{q_2^\prime}}{(2\pi)^32E_{q_2^\prime}}
         \frac {d^3p_{\bar {q}_2^\prime}}{(2\pi)^32E_{\bar {q}_2^\prime}}
                    \nonumber   \\
& & \mid {\cal M}_{\rm fi} \mid^2 \delta (P_A + P_B - p_{q_1^\prime} 
-p_{\bar {q}_1^\prime}  - p_{q_2^\prime} - p_{\bar {q}_2^\prime} )      
                    \nonumber   \\
& = & \frac {1}{32(2\pi)^8\sqrt {[s-(m_A+m_B)^2]
[s-(m_A-m_B)^2]}} \int d\Omega_{q_2^\prime}
\frac {d^3p_{q_1^\prime}}{E_{q_1^\prime}}
\frac {d^3p_{\bar {q}_1^\prime}}{E_{\bar {q}_1^\prime}} 
                              \nonumber   \\
& & \frac {\vec {p}_{q_2^\prime}^2 \mid {\cal M}_{\rm fi} \mid^2}
{\mid \mid \vec {p}_{q_2^\prime} \mid E_{\bar {q}_2^\prime}
+(\mid \vec {p}_{q_2^\prime} \mid -
\mid \vec {P}_A + \vec {P}_B - \vec {p}_{q_1^\prime}
 - \vec {p}_{\bar {q}_1^\prime} \mid \cos \Theta ) E_{q_2^\prime} \mid}
                              \nonumber   \\
\end{eqnarray}
where $m_A$ ($m_B$) and $P_A=(E_A,\vec {P}_A)$ ($P_B=(E_B,\vec {P}_B)$) are
the mass and the four-momentum of meson $A$ ($B$), respectively, and 
$s=(P_A+P_B)^2$; $p_i=(E_i,\vec {p}_i)$ ($i=q_1^\prime, \bar {q}_1^\prime, 
q_2^\prime, \bar {q}_2^\prime$) is the 
four-momentum of a final quark or antiquark, and subscripts of variables for
the final quarks and antiquarks are labeled with primes;
$\Theta$ is the angle between
$\vec {p}_{q_2^\prime}$ and $\vec {P}_A + \vec {P}_B - \vec {p}_{q_1^\prime}
 - \vec {p}_{\bar {q}_1^\prime}$, and $d\Omega_{q_2^\prime}$ is the solid 
angle centered about the direction of $\vec {p}_{q_2^\prime}$.
The transition amplitude ${\cal M}_{\rm fi}$ arising from one gluon exchange
between one constituent in meson $A$ and one constituent in meson $B$ is
\begin{eqnarray}
{\cal M}_{\rm fi} & = & \sqrt {2E_A2E_B
2E_{q_1^\prime}2E_{\bar {q}_1^\prime}2E_{q_2^\prime}2E_{\bar {q}_2^\prime}}
                      \nonumber   \\
& & \left[ V_{q_1\bar {q}_2}(\vec {Q})
\psi_{q_1\bar {q}_1} (\vec {p}_{q_1^\prime \bar {q}_1^\prime}
-\frac {m_{\bar {q}_1}}{m_{q_1}+m_{\bar {q}_1}}\vec {Q}) 
\psi_{q_2\bar {q}_2} (\vec {p}_{q_2^\prime \bar {q}_2^\prime}
-\frac {m_{q_2}}{m_{q_2}+m_{\bar {q}_2}}\vec {Q})     \right.
                      \nonumber \\
& &  + V_{\bar {q}_1q_2}(\vec {Q})
\psi_{q_1\bar {q}_1} (\vec {p}_{q_1^\prime \bar {q}_1^\prime}
+\frac {m_{q_1}}{m_{q_1}+m_{\bar {q}_1}}\vec {Q}) 
\psi_{q_2\bar {q}_2} (\vec {p}_{q_2^\prime \bar {q}_2^\prime}
+\frac {m_{\bar {q}_2}}{m_{q_2}+m_{\bar {q}_2}}\vec {Q}) 
                      \nonumber \\
& &  + V_{q_1q_2}(\vec {Q})
\psi_{q_1\bar {q}_1} (\vec {p}_{q_1^\prime \bar {q}_1^\prime}
-\frac {m_{\bar {q}_1}}{m_{q_1}+m_{\bar {q}_1}}\vec {Q}) 
\psi_{q_2\bar {q}_2} (\vec {p}_{q_2^\prime \bar {q}_2^\prime}
+\frac {m_{\bar {q}_2}}{m_{q_2}+m_{\bar {q}_2}}\vec {Q}) 
                      \nonumber \\
& &  \left.  + V_{\bar {q}_1\bar{q}_2}(\vec {Q})
\psi_{q_1\bar {q}_1} (\vec {p}_{q_1^\prime \bar {q}_1^\prime}
+\frac {m_{q_1}}{m_{q_1}+m_{\bar {q}_1}}\vec {Q}) 
\psi_{q_2\bar {q}_2} (\vec {p}_{q_2^\prime \bar {q}_2^\prime}
-\frac {m_{q_2}}{m_{q_2}+m_{\bar {q}_2}}\vec {Q})   \right]
                      \nonumber \\
\end{eqnarray}
where $\vec Q$ is the gluon momentum, $m_i$ ($i=q_1, \bar {q}_1, q_2, \bar 
{q}_2$) is the mass of a constituent quark or antiquark of mesons, and 
$\vec {p}_{ij}$ is the relative momentum of quark $i$ and antiquark $j$.
For mesons $\psi_{ij} (\vec {p}_{ij})$ is the wave function of the 
quark-antiquark relative motion in momentum space and satisfies
$\int \frac {d^3p_{ij}}{(2\pi)^3} \psi^+_{ij} (\vec {p}_{ij})
\psi_{ij} (\vec {p}_{ij}) =1$. 
$\mathcal {M}_{\rm fi}$ is deriveded from the matrix element
\begin{equation}
\langle q_1, \bar {q}_1, q_2, \bar {q}_2 \mid H_I
\mid A, B \rangle 
= \frac {(2\pi)^3 \delta (\vec {P}_{\rm i} - \vec {P}_{\rm f} )
\mathcal {M}_{\rm fi} } {V^3 \sqrt {2E_A2E_B
2E_{q_1^\prime}2E_{\bar {q}_1^\prime}2E_{q_2^\prime}2E_{\bar {q}_2^\prime} }}
\end{equation}
where $\vec {P}_{\rm i}$ ($\vec {P}_{\rm f}$) is the total three-dimensional 
momentum of the two initial mesons (final quarks and antiquarks);
the wave functions of the initial mesons and of the final quarks and
antiquarks are normalized to one in volume $V$, respectively. The interaction
is
\begin{equation}
H_{\rm I}= \tilde {V}(\vec {r}_{q_1\bar {q}_2})
  +\tilde {V}(\vec {r}_{\bar {q}_1 q_2})
  +\tilde {V}(\vec {r}_{q_1q_2}) +\tilde {V}(\vec {r}_{\bar {q}_1 \bar {q}_2}) 
\end{equation}
where $\tilde {V}(\vec {r}_{ij})$ is the potential 
and $\vec {r}_{ij}$ is the relative coordinate of constituents $i$ and $j$. 
The potential takes a form \cite {wong} whose central interquark potential is 
obtained from the lattice gauge results of Karsch {\it et al.} \cite {KLP}. 
The potential reflects the medium screening effect at finite temperature.
It has been used to calculate temperature-dependent dissociation cross
sections of $\pi + J/\psi$, $\pi + \chi_{c1}$ and $\pi + \chi_{c2}$ \cite {XWB}
in the quark-interchange mechanism \cite {BS92}. 
Fourier transform of the coordinate-space potential leads to the
expression in momentum space
\begin{equation}
V_{ij}(\vec {Q})= \frac {\vec {\lambda}_i}{2} \cdot \frac {\vec {\lambda}_j}{2}
\left[   \frac {4\pi \alpha_s }{\mu^2 (T)+\vec {Q}^2}
+ \frac {6\pi b(T)}{(\mu^2 (T)+\vec {Q}^2)^2}
- \frac {8\pi \alpha_s}{3m_im_j} \vec {s}_i \cdot \vec {s}_j
\exp \left( - \frac {\vec {Q}^2}{4d^2} \right)    \right]
\end{equation}
where $\vec {\lambda}_i$, $\vec {s}_i$ and $m_i$ are the Gell-Mann 
"$\lambda$-matrices", the spin and the mass of the constituent $i$, 
respectively; $d=0.897$ GeV \cite {WSB,BSWX}, $\alpha_s = \frac {12\pi}
{25 \ln (10+Q^2/X^2)}$ with $X=0.31$ GeV,
$b(T)=b_0[1-(T/T_c)^2]\theta (T_c-T)$ with $b_0 =0.35$ ${\rm GeV}^2$ and 
the critical temperature $T_c=0.175$ GeV, and $\mu (T) =\mu_0 \theta (T_c-T)$ 
with $\mu_0=0.28$ GeV. Determined by meson spectroscopy, the constituent quark
masses (CQM) are 0.334 GeV for both the up quark and the down quark and 0.575 
GeV for the strange quark \cite {WSB}.

Denote the orbital angular momentum and the spin of meson $A$ by $L_A$ and 
$S_A$, respectively. Similar notation is established for meson $B$. For the 
three cases: (1) $L_A=L_B=0$, (2) $L_A=0$, $L_B \ne 0$, $S_A=0$, (3) $L_A=0$, 
$L_B=1$, $S_A=1$, $S_B=1$, the unpolarized cross section is
\begin{eqnarray}
\sigma ^{\rm unpol}_{AB \to {\rm free}} (\sqrt {s}) & = &
\frac{1}{(2S_A+1)(2S_B+1)(2L_B+1)}                 \nonumber   \\
& &   \sum _{L_{Bz}SS_{q_1^\prime+\bar {q}_1^\prime}
                    S_{q_2^\prime+\bar {q}_2^\prime}} (2S+1)
\sigma(L_{Bz},S,m_S,S_{q_1^\prime+\bar {q}_1^\prime},
                    S_{q_2^\prime+\bar {q}_2^\prime},\sqrt {s})
                           \nonumber    \\
\end{eqnarray}
where $L_{Bz}$ is the magnetic quantum number of meson $B$, the total spin
$S$ takes values that are allowed by $\mid S_A-S_B \mid \leq S \leq S_A +S_B$
and $m_S$ is the component of $S$.
$S_{q_1^\prime+\bar {q}_1^\prime}$ ($S_{q_2^\prime+\bar {q}_2^\prime}$) is the
total spin of the final constituents $q_1$ and $\bar {q}_1$ 
($q_2$ and $\bar {q}_2$).
The spins $S$, $S_{q_1^\prime+\bar {q}_1^\prime}$ and 
$S_{q_2^\prime+\bar {q}_2^\prime}$ satisfy
$\mid S_{q_1^\prime+\bar {q}_1^\prime} 
-S_{q_2^\prime+\bar {q}_2^\prime} \mid \leq S \leq 
S_{q_1^\prime+\bar {q}_1^\prime} +S_{q_2^\prime+\bar {q}_2^\prime}$.
$\sigma$ is independent of $m_S$ and is calculated at any value subject to the
condition $-S \leq m_S \leq S$.

The Schr$\rm \ddot o$dinger equation with the temperature-dependent potential
$\tilde {V}(\vec {r}_{ij})$ \cite{wong}
produces temperature-dependent quark-antiquark wave functions
which Fourier transform produces momentum-space wave functions used in 
obtaining the transition amplitude. Cross sections depend on temperature as
well as the center-of-mass energy of the two initial mesons. While temperature
increases, the confinement potential gets weak and the bound state gets loose.
At higher temperature stronger screening leads to larger cross sections for
meson-meson reactions.

To uniquely show the role of $A(q_1\bar {q}_1)+B(q_2\bar {q}_2) 
\to q_1+\bar {q}_1+q_2 +\bar {q}_2$, we do not consider any expansion of 
hadronic matter. The master rate equations for free quarks, pions, rhos, 
kaons and vector kaons in static hadronic matter are
\begin{equation}
\frac {dn_q}{dt} = \sum_{i=\pi,\rho,K,K^\ast} \sum_{j=\pi,\rho,K,K^\ast}
\langle v_{\rm rel}\sigma^{\rm unpol}_{ij\to {\rm free}}\rangle n_i n_j
\end{equation}
\begin{equation}
\frac {dn_\pi}{dt} = -\sum_{j=\pi,\rho,K,K^\ast}
\langle v_{\rm rel}\sigma^{\rm unpol}_{\pi j\to {\rm free}}\rangle n_\pi n_j
\end{equation}
\begin{equation}
\frac {dn_\rho}{dt} = -\sum_{j=\pi,\rho,K,K^\ast}
\langle v_{\rm rel}\sigma^{\rm unpol}_{\rho j\to {\rm free}}\rangle n_\rho n_j
\end{equation}
\begin{equation}
\frac {dn_K}{dt} = -\sum_{j=\pi,\rho,K,K^\ast}
\langle v_{\rm rel}\sigma^{\rm unpol}_{K j\to {\rm free}}\rangle n_K n_j
\end{equation}
\begin{equation}
\frac {dn_{K^\ast}}{dt} = -\sum_{j=\pi,\rho,K,K^\ast}
\langle v_{\rm rel}\sigma^{\rm unpol}_{K^\ast j\to {\rm free}}\rangle 
n_{K^\ast} n_j
\end{equation}
where $n_q$, $n_\pi$, $n_\rho$, $n_K$ and $n_{K^\ast}$ are the number densities
of free quarks, $\pi$, $\rho$, $K$ and $K^\ast$, respectively; 
$v_{\rm rel}$ in the thermal averages denoted by the symbols $< \cdot \cdot 
\cdot >$ is the relative velocity of two colliding hadrons. The number density 
of free antiquarks equals the one of free quarks.

The master rate equations are solved from pure hadronic matter in thermal
equilibrium at $t=0$ fm/$c$. The initial number densities of free quarks and 
antiquarks equal zero, respectively,
and the number densities of $\pi$, $\rho$, $K$ and $K^\ast$ are given by
$n_j = \int \frac {d^3p_j}{(2\pi)^3} \frac {g_j}{\exp 
(\sqrt {\vec {p}_j^2+m_j^2}/T)-1}$
with the degeneracy factor $g_j$, the hadron mass $m_j$ and $j=\pi, \rho, K, 
K^\ast$. The equation solutions as functions of time at different temperatures
are shown in Figs. 1-3. When more and more hadrons due to the collisions 
convert
into free quarks and antiquarks, the number density of free quarks increases 
with increasing time and accordingly the number density of each hadron species 
decreases. The number densities vary faster at higher temperature because of
larger cross sections.
We present two solutions for $T=0.174$ GeV near the critical temperature. The
solution shown by the dashed curves is obtained for the final quarks and 
antiquarks taking the constituent quark 
masses. The other solution shown by the solid curves corresponds to the chiral
limit (CL) of final quarks and antiquarks. A comparison of the five dashed
curves shows that 
the number density of free quarks exceeds the number densities
of $\pi$, $\rho$, $K$ and $K^\ast$ at $t=0.83$ fm/$c$, 0.21 fm/$c$,  
0.33 fm/$c$ and 0.18 fm/$c$, respectively. In the chiral limit
the number density of free quarks exceeds the number densities of $\pi$, 
$\rho$, $K$ and $K^\ast$ at $t=0.28$ fm/$c$, 0.1 fm/$c$, 0.13 fm/$c$ and 
0.08 fm/$c$, respectively. Therefore, while the chiral symmetry is restored,
hadrons convert quickly into free quarks and antiquarks.

\begin{table}[htbp]
\centering \caption{Ratios of number densities at $t=0.5$ fm/$c$.}
\label{ratios}
\begin{tabular*}{16cm}{@{\extracolsep{\fill}}cccccc}
  \hline
  $T~({\rm GeV})$ & $n_q/n_\pi$ & $n_q/n_\rho$ & $n_q/n_K$ & $n_q/n_{K^\ast}$ 
& $n_q/(n_\pi+n_\rho+n_K+n_{K^\ast})$  \\
  \hline
  0.174 (CL) & 1.83 & 5.71 & 3.75 & 6.33 & 0.87\\
  0.174 (CQM) & 0.63 & 2.60 & 1.50 & 2.86 & 0.33\\
  0.16 & 0.19 & 1.01 & 0.53 & 1.23 & 0.11 \\
  0.15 & 0.08 & 0.46 & 0.23 & 0.62 & 0.05 \\
  0.14 & 0.03 & 0.20 & 0.10 & 0.29 & 0.02 \\
  \hline
\end{tabular*}
\end{table}

The ratio of the free-quark number density to the pion number density etc.
are listed in Table 1. The number density of hadronic matter is approximately
equal to $n_\pi+n_\rho+n_K+n_{K^\ast}$.
The first and second rows are obtained in the
chiral limit and in the use of the constituent quark masses, respectively.
In such a short time period of 0.5 fm/$c$ a large amount of quarks are
produced near the critical temperature by $A(q_1\bar {q}_1)+B(q_2\bar {q}_2) 
\to q_1+\bar {q}_1+q_2 +\bar {q}_2$ so that the free quarks are very important 
in hadronic matter. At $T=0.16$ GeV an appreciable amount of free quarks exist
in hadronic matter. When temperature is lower than 0.14 GeV, the ratio 
$n_q/(n_\pi+n_\rho+n_K+n_{K^\ast})$ is 
so small that free quarks can be neglected. This can also be recognized by
the flat long dashed lines in Figs. 1-3.

At $T=0.174$ GeV, if all of $\pi$, $\rho$, $K$ and $K^\ast$ break up
into free quarks and antiquarks, the number density of free quarks is 0.40
fm$^{-3}$. In the chiral limit about 75\% of the four species of hadrons 
dissociate at $t=1.7$ fm/$c$ and 90\% at $t=4.7$ fm/$c$. While the constituent
quark masses are used, about 70\% of the hadrons dissociate at $t=5.0$
fm/$c$ and 90\% at $t=16.0$ fm/$c$. We have seen now that the deconfinement
process of hadronic matter in the chiral limit is much faster than in
the use of the constituent quark masses. 

As a first step that we study the role of $A(q_1\bar {q}_1)+B(q_2\bar {q}_2) 
\to q_1+\bar {q}_1+q_2 +\bar {q}_2$, we haven't considered quark-antiquark
annihilation process.
If the contribution of the annihilation process was included, more
quarks and antiquarks would be produced. 

In conclusion, the reaction $A(q_1\bar {q}_1)+B(q_2\bar {q}_2) 
\to q_1+\bar {q}_1+q_2 +\bar {q}_2$ gets important in hadronic matter at high
temperature because of the medium screening effect. Due to the reactions of 
$\pi$, $\rho$, $K$ and $K^\ast$, hadronic matter is mixed with free quarks and
antiquarks. The reaction is a new dynamical process for the deconfinement 
transition of hadronic matter.

\vspace{0.5cm}
\leftline{\bf Acknowledgements}
\vspace{0.5cm}
We thank colleagues for having read the manuscript.
This work was supported by National Natural Science Foundation of China under 
Grant No. 10675079.

\newpage

\begin{figure}[htbp]
\centering
\includegraphics[width=12cm]{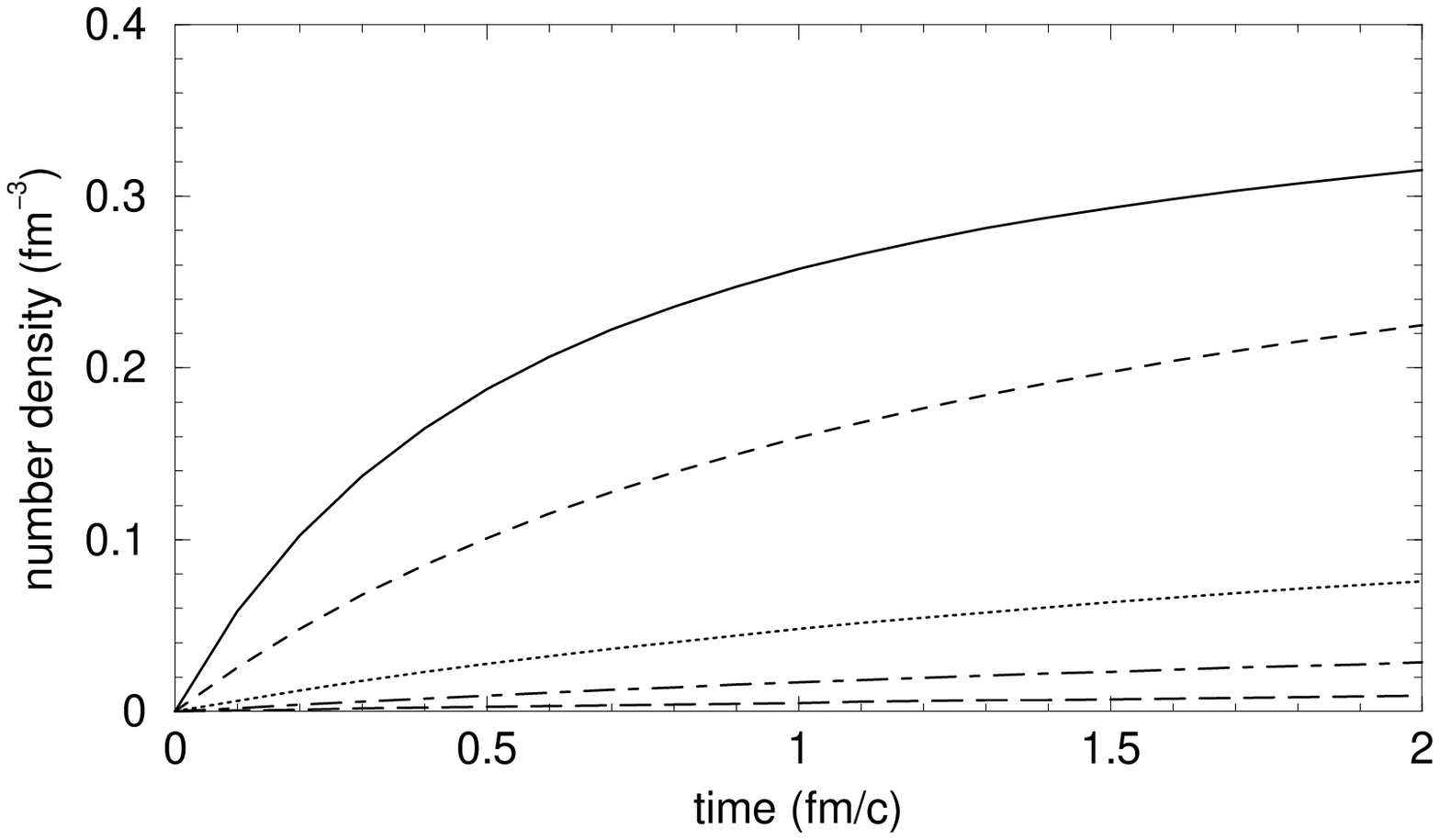}%
\caption{Number densities of free quarks or antiquarks with the constituent 
quark masses at $T=0.14$ GeV (long dashed), 0.15 GeV (dot-dashed), 0.16 
(dotted), 0.174 GeV (dashed). The solid curve denotes the number density in the
chiral limit at $T=0.174$ GeV.}
\label{fig1}
\end{figure}

\begin{figure}[htbp]
\centering
\includegraphics[scale=0.7]{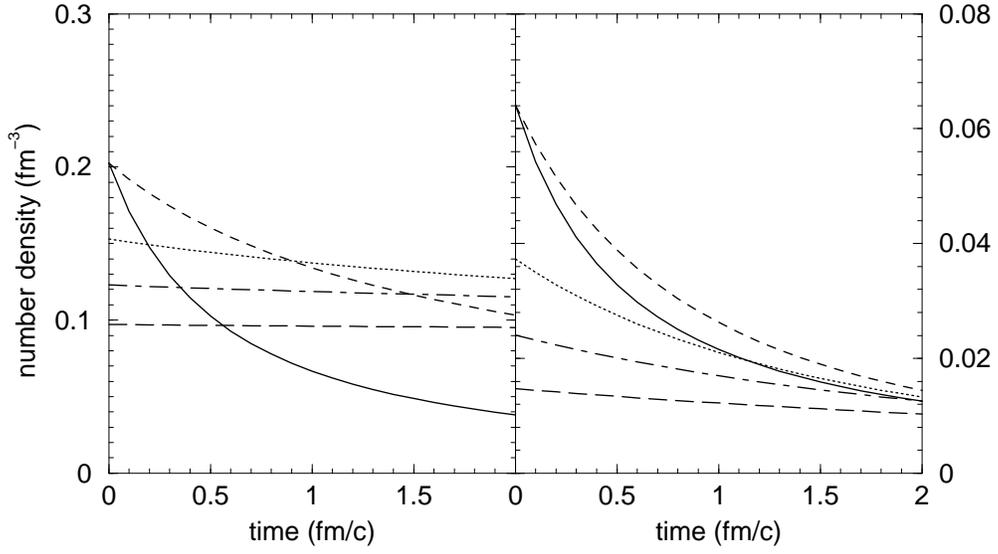}%
\caption{The same as Fig. 1 except for pions in the left panel and rhos in
the right panel.}
\label{fig2}
\end{figure}

\begin{figure}[htbp]
\centering
\includegraphics[scale=0.7]{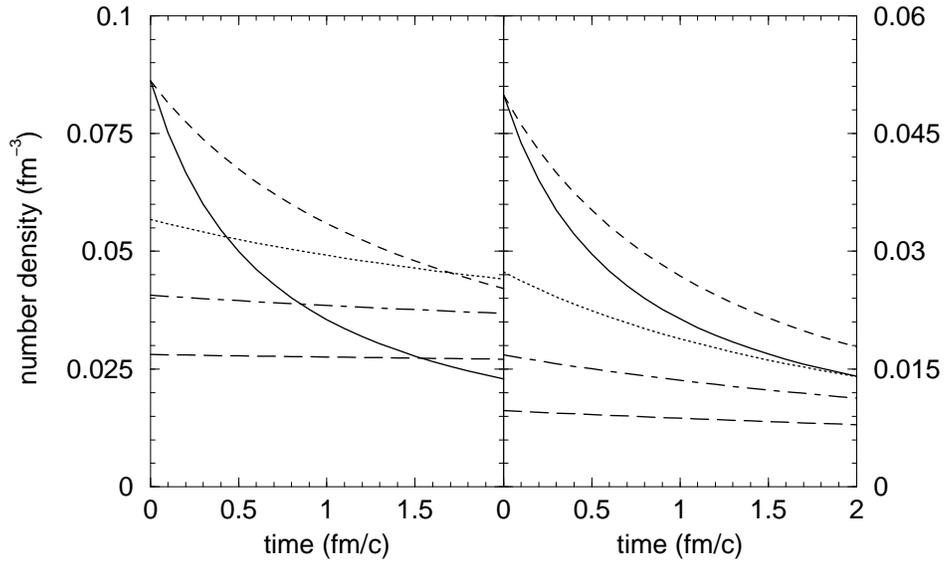}%
\caption{The same as Fig. 1 except for kaons in the left panel and vector
kaons in the right panel.}
\label{fig3}
\end{figure}

\end{document}